# A Formally Specified Program Logic for Higher-Order Procedural Variables and non-local Jumps

**Tristan Crolard**

*December 2011*

**Laboratoire d'Algorithmique, Complexité et Logique (LACL)**
**Département d'Informatique**
**Université Paris-Est Créteil – Val de Marne, Faculté des Sciences et Technologie**
61, Avenue du Général de Gaulle, 94010 Créteil cedex, France
Tel.: (33)(1) 45 17 16 47, Fax: (33)(1) 45 17 66 01

Laboratory of Algorithmics, Complexity and Logic (LACL)
University Paris-Est Créteil

Tristan Crolard.

*A Formally Specified Program Logic for Higher-Order Procedural Variables and non-local Jumps*

# A Formally Specified Program Logic for Higher-Order Procedural Variables and non-local Jumps

T. Crolard

December, 2011

**Abstract**

We formally specified a program logic for higher-order procedural variables and non-local jumps with Ott and Twelf. Moreover, the dependent type systems and the translation are both executable specifications thanks to Twelf's logic programming engine. In particular, relying on Filinski's encoding of **shift/reset** using **callcc/throw** and a global meta-continuation (simulated in state passing style), we have mechanically checked the correctness of a few examples (all source files are available on request).

## 1 Introduction

We formally specified the formal systems described in [Cro10, CP11] with Ott [SNO$^+$07] and the Twelf proof assistant [PS99]. These formal systems are:

- The functional language **F** (which is our formulation of Gödel System T) equipped with two usual type systems, a simple type system **IS** and a dependent type system **ID** which is akin to Leivant's **M1LP** [Lei90]. In particular, dependent types include arbitrary formulas of first-order arithmetic.
- The imperative language **I** (essentially LOOP$^\omega$ from [CPV09]) is an extension of Meyer and Ritchie's Loop language [MR76] with higher-order procedural variables. Language **I** is also equipped with two (unusual) type systems, a pseudo-dynamic simple type system **IS** and a dependent type system **ID**.
- A compositional translation from **I** to **F** is also defined [CPV09] in both the pseudo-dynamic and dependent frameworks.

The main difference from the description given in [CP11] comes from the fact that the dependently-typed programs contain proof annotations and are actually isomorphic to proof derivations (this is required to obtain executable proof checkers from the specification of the dependent type systems in Twelf). As a simple example of such proof annotations, the dependently-typed imperative procedure for addition is given in Figure 1.

A second minor difference is a consequence of our encoding of first-order quantifiers using Twelf higher-order abstract syntax. Quantified variables have to be dealt with separately, and the elimination rule for the existential quantifier is thus split into a cut rule and a left introduction rule.

Moreover, the type systems and the translation are all executable specifications thanks to Twelf's logic programming engine. In particular, the imperative counterpart of Filinski's encoding of shift/reset [DF89, Fil94] described in [CP11] and the examples from [Wad94] have been mechanically checked. The correctness of third example (which requires the more general type system) is shown in full in Figure 2.

In Section 2, we present syntax of **I** and **F**, the functional simple type system **FS** (Section 2.1), the imperative pseudo-dynamic type system **IS** (Section 2.2) and the translation form **IS** to **FS** (Section 2.3). In Section 3, we present syntax of languages **I** and **F** extended with dependent types and proof annotations, the functional dependent type system **FS** (Section 3.1), the pseudo-dynamic imperative dependent type system **ID** (Section 3.2) and the translation form **ID** to **FD** (Section 3.3).

$\mathbf{cst}\ p_add = \mathbf{proc}\ \forall n \forall m [x{:}\mathbf{nat}(n),\ y{:}\mathbf{nat}(m)]\ \mathbf{out}\ [z{:}\mathbf{nat}(add(n,m))]\ \{$
$\quad z := y :> \{i/\mathbf{nat}(i)\}\ [add(0,\ m) = m];$
$\quad \mathbf{for}\ l{:}\mathbf{nat}(l) := 0\ \mathbf{until}\ x\ \{$
$\quad\quad \mathbf{inc}(z);$
$\quad\quad z := z :> \{i/\mathbf{nat}(i)\}\ [add(succ(l),m) = succ(add(l,m))];$
$\quad \}z{:}\mathbf{nat}(add(l,m));$
$\};$

Figure 1: Dependently-typed addition

```
cst shift = proc [p:proc ([proc ∀n([nat(n), ∼A] out [nat(F_32(n)), ∼A]), ∼proc ∀n([nat(n), ∼A] out [nat(F_32(n)), ∼A])]
                          out [proc ∀n([nat(n), ∼A] out [nat(F_32(n)), ∼A]), ∼proc ∀n([nat(n), ∼A] out [nat(F_32(n)), ∼A])]),
                 mk2:∼proc ∀n([nat(n), ∼A] out [nat(F_32(n)), ∼A])] out ∃u[r:nat(u), mk:∼nat(F_32(u))] {
    mk := mk2;
    cst reset = proc ∀x[p:proc ([∼nat(F_32(x))] out [H, ∼H]), mk2:∼A] out [r:nat(F_32(x)), mk:∼A] {
        mk := mk2;
        k :{
            cst m = mk;
            mk := proc [r:nat(F_32(x))] out [Z:⊥] {
                jump(k, r, m)[Z:⊥];
            };
            var y := ∗;
            p(mk; y, mk);
            jump(mk, y)[r:nat(F_32(x)), mk:∼A];
        }[r:nat(F_32(x)), mk:∼A];
    };
    k :{
        cst q = proc ∀x[v:nat(x), mk2:∼A] out [r:nat(F_32(x)), mk:∼A] {
            mk := mk2;
            cst anonym = proc [mk2:∼nat(F_32(x))] out [z:H, mk:∼H] {
                mk := mk2;
                jump(k <: {u/[nat(u), ∼nat(F_32(u))]}{x}, v, mk)[z:H, mk:∼H];
            };
            reset{x}(anonym, mk; r, mk);
        };
        var y := ∗;
        p(q, mk; y, mk);
        jump(mk, y)[r:nat(0), mk:∼nat(F_32(0))];
        [ 0 ∈ ∃u[r:nat(u), mk:∼nat(F_32(u))] ]
    }∃u[r:nat(u), mk:∼nat(F_32(u))];?u.
    [ u ∈ ∃u[r:nat(u), mk:∼nat(F_32(u))] ]
};
cst reset = proc [p:proc ([∼proc ∀n([nat(n), ∼A] out [nat(F_32(n)), ∼A])] out ∃v[nat(v), ∼nat(v)]), mk2:∼A]
                                   out [r:proc ∀n([nat(n), ∼A] out [nat(F_32(n)), ∼A]), mk:∼A] {
    mk := mk2;
    k :{
        cst m = mk;
        mk := proc [r:proc ∀n([nat(n), ∼A] out [nat(F_32(n)), ∼A])] out [Z:⊥] {
            jump(k, r, m)[Z:⊥];
        };
        var y := ∗;
        p(mk; y, mk);?v.
        jump(mk, y)[r:proc ∀n([nat(n), ∼A] out [nat(F_32(n)), ∼A]), mk:∼A];
    }[r:proc ∀n([nat(n), ∼A] out [nat(F_32(n)), ∼A]), mk:∼A];
};
cst a = proc [mk2:∼A] out [z:nat(add(3, 2)), mk:∼A] {
    cst p_add = proc {x}∀y[X:nat(x), Y:nat(y), mk2:∼A] out [Z:nat(add(x, y)), mk:∼A] {
        mk := mk2;
        Z := X :> {var_2/nat(var_2)}[add(x, 0) = x];
        for i : nat(i) := 0 until Y {
            inc(Z);
            ( :> {var_3}[Z:nat(var_3)][add(x, succ(i)) = succ(add(x, i))])
        }[Z:nat(add(x, i))];
    };
    cst q = proc [mk2:∼proc ∀n([nat(n), ∼A] out [nat(F_32(n)), ∼A])] out ∃v[r:nat(v), mk:∼nat(v)] {
        mk := mk2;
        cst p = proc [f:proc ∀n([nat(n), ∼A] out [nat(F_32(n)), ∼A]), mk2:∼proc ∀n([nat(n), ∼A] out [nat(F_32(n)), ∼A])]
                     out [h:proc ∀n([nat(n), ∼A] out [nat(F_32(n)), ∼A]), mk:∼proc ∀n([nat(n), ∼A] out [nat(F_32(n)), ∼A])] {
            mk := mk2;
            h := f;
        };
        var b := ∗;
        shift(p, mk; b, mk);?u.
        r := 3 :> {var_4/nat(var_4)}[F_32(0) = 3];
        for i : nat(i) := 0 until b {
            r := 2 :> {var_5/nat(var_5)}[F_32(succ(i)) = 2];
        }[r:nat(F_32(i))];
        [ F_32(u) ∈ ∃v[r:nat(v), mk:∼nat(v)] ]
    };
    var mk := mk2;
    var g := ∗;
    reset(q, mk; g, mk);
    var x := ∗;
    g{0}(0, mk; x, mk);
    var y := ∗;
    g{1}(1, mk; y, mk);
    p_add{3}{2}(x :> {var_6/nat(var_6)}[3 = F_32(0)], y :> {var_7/nat(var_7)}[2 = F_32(1)], mk; z, mk);
};
```

Figure 2: Dependently-typed example with shift/reset (imperative version of example 3 from [Wad94])

## 2 Grammars and judgments for FS and IS

| | | | | |
|---|---|---|---|---|
| *ident*, $x$, $y$, $z$ | ::= | | | variable: |
| | \| | *variable* | | |
| | | | | |
| *idents*, $\vec{x}$, $\vec{y}$, $\vec{z}$ | ::= | | | variables: |
| | \| | | | |
| | \| | $\$$ | S | |
| | \| | $\vec{x}, x$ | | |
| | \| | $x$ | S | |
| | \| | $(\vec{x})$ | S | |
| | | | | |
| *fenv*, $\Sigma$ | ::= | | | Environments: |
| | \| | $\{\}$ | | empty environment |
| | \| | $\#$ | S | |
| | \| | $\Sigma, x : \tau$ | | ident declaration |
| | | | | |
| *terms*, $\vec{t}$, $\vec{u}$ | ::= | | | Variables: |
| | \| | | S | |
| | \| | $\vec{t}, t$ | S | |
| | \| | $t$ | S | |
| | \| | $(\vec{t})$ | S | |
| | | | | |
| *term*, $t$, $u$ | ::= | | | Term: |
| | \| | $x$ | | var |
| | \| | $0$ | | zero |
| | \| | $t_1\ t_2$ | | application |
| | \| | $\mathbf{fn}\, x : \tau \Rightarrow t$ | | abstraction |
| | \| | $\mathbf{fn}\, (\vec{x} : \vec{\tau}) \Rightarrow t$ | | multi-abstraction |
| | \| | $\mathbf{succ}\, (t)$ | | successor |
| | \| | $\mathbf{pred}\, (t)$ | | predecessor |
| | \| | $\mathbf{rec}\, (t_1, t_2, t_3)$ | | recursor |
| | \| | $\mathbf{let}\, x = t_1\, \mathbf{in}\, t_2$ | | let |
| | \| | $\mathbf{let}\, \langle \vec{x} \rangle = t_1\, \mathbf{in}\, t_2$ | | match |
| | \| | $\langle \vec{t} \rangle$ | | tuple |
| | \| | $(t)$ | S | |
| | | | | |
| *typ*, $\tau$ | ::= | | | Type: |
| | \| | $\top$ | | unit |
| | \| | $\bot$ | | void |
| | \| | $\mathbf{nat}$ | | nat |
| | \| | $\tau \to \tau'$ | | imply |
| | \| | $\sim \tau$ | | not |
| | \| | $\langle \vec{\tau} \rangle$ | | tuple |
| | \| | $(\tau)$ | S | |
| | | | | |
| *typs*, $\vec{\tau}$ | ::= | | | Types: |
| | \| | | S | |
| | \| | $\vec{\tau}, \tau$ | S | |
| | \| | $\tau$ | S | |
| | \| | $(\vec{\tau})$ | S | |
| | | | | |
| *env*, $\Gamma$, $\Omega$, $\gamma$, $\omega$ | ::= | | | Environments: |
| | \| | | | empty environment |

| | | | | |
|---|---|---|---|---|
| | \| | $\Gamma, x : \tau$ | | ident declaration |
| | \| | $x : \tau$ | S | ident declaration |
| | \| | $(\Gamma)$ | S | |
| $b$ | ::= | | | block: |
| | \| | $\{s\}_\omega$ | | block |
| *command*, $c$ | ::= | | | command: |
| | \| | $b$ | | block |
| | \| | $\textbf{for}\, y := 0\, \textbf{until}\, e\, b$ | | for |
| | \| | $y := e$ | | assign |
| | \| | $\textbf{inc}\,(y)$ | | inc |
| | \| | $\textbf{dec}\,(y)$ | | dec |
| | \| | $e(\vec{e}; \vec{y})$ | | call |
| *sequence*, $s$ | ::= | | | sequence: |
| | \| | | | empty sequence |
| | \| | $\varepsilon$ | S | empty sequence |
| | \| | $c; s$ | | command |
| | \| | $\textbf{cst}\, y = e; s$ | | constant |
| | \| | $\textbf{var}\, y := e; s$ | | variable |
| | \| | $\textbf{var}\, y; s$ | S | variable |
| | \| | $(s)$ | S | |
| *number*, $q$ | ::= | | | number: |
| | \| | $0$ | | zero |
| | \| | $1$ | S | |
| | \| | $2$ | S | |
| | \| | $3$ | S | |
| | \| | $4$ | S | |
| | \| | $5$ | S | |
| | \| | $\textbf{succ}\,(q)$ | | successor |
| *expression*, $e$, $p$ | ::= | | | expression: |
| | \| | $x$ | | variable |
| | \| | $\star$ | | star |
| | \| | $q$ | | number |
| | \| | $\textbf{proc}\,[\gamma]\,\textbf{out}\,[\omega]\{s\}$ | | procedure |
| *expressions*, $\vec{e}$ | ::= | | | expressions: |
| | \| | | | |
| | \| | $\vec{e}, e$ | | |
| | \| | $e$ | S | |
| | \| | $(\vec{e})$ | S | |
| *prop*, $\tau$, $\sigma$ | ::= | | | proposition: |
| | \| | $\top$ | | unit |
| | \| | $\textbf{nat}$ | | nat |
| | \| | $\textbf{proc}\,([\vec{\tau}]\,\textbf{out}\,[\vec{\tau}'])$ | | proc |
| | \| | $(\tau)$ | S | |
| *props*, $\vec{\tau}$, $\vec{\sigma}$ | ::= | | | propositions: |
| | \| | | | |
| | \| | $\vec{\tau}, \tau$ | | |
| | \| | $\tau$ | S | |
| | \| | $(\vec{\tau})$ | S | |
| *primitives* | ::= | | | |

| | | | |
|---|---|---|---|
| *f_typing* | ::= | | |
| | \| | $\tau = \tau'$ | Formulas equality |
| | \| | $t = t'$ | Terms equality |
| | \| | $x : \tau \in \Sigma$ | Lookup |
| | \| | $\Sigma \vdash t : \tau$ | Type check |
| | \| | $\Sigma, \vec{x} : \vec{\tau} = \Sigma'$ | Append |
| | \| | $\Sigma, \langle \vec{x} \rangle : \tau \vdash t : \tau'$ | Type check term in extended environment |
| | \| | $\Sigma \vdash (\vec{t}) : (\vec{\tau})$ | Type check terms |

| | | | |
|---|---|---|---|
| *typing* | ::= | | |
| | \| | $\tau = \tau'$ | Propositions equality |
| | \| | $x : \tau \in \Gamma$ | Lookup ident |
| | \| | $\vec{x} : \vec{\tau} \subset \Gamma$ | Lookup idents |
| | \| | $\Omega[x : \tau] = \Omega'$ | Update |
| | \| | $\Omega[\![\vec{x} : \vec{\tau}]\!] = \Omega'$ | Multi-update |
| | \| | $\Gamma, \gamma = \Gamma'$ | Append |
| | \| | $\omega \subset \Omega$ | Subset |
| | \| | $\Omega_{\|\vec{x}} = \omega$ | Restriction |
| | \| | $\Omega = \vec{x} : \vec{\tau}$ | Split |
| | \| | $\vec{x} : \vec{\top} = \omega$ | Init |
| | \| | $\Gamma; \Omega \vdash e : \tau$ | Typecheck expression |
| | \| | $\Gamma; \Omega \vdash (\vec{e}) : (\vec{\tau})$ | Typecheck expressions |
| | \| | $\Gamma; \Omega \vdash s \triangleright \Omega'$ | Typecheck sequence |

| | | | |
|---|---|---|---|
| *translation* | ::= | | |
| | \| | $(\tau)^\star = \tau$ | Types translation |
| | \| | $(\vec{\tau})^\star = (\vec{\tau})$ | Types translation |
| | \| | $(\vec{x})^\star = \vec{t}$ | Sequence translation |
| | \| | $q^\star = t$ | Number translation |
| | \| | $(e)^\star = t$ | Expression translation |
| | \| | $(\vec{e})^\star = \vec{t}$ | Expressions translation |
| | \| | $(s)^\star_{\vec{x}} = t$ | Sequence translation |

| | | |
|---|---|---|
| *judgement* | ::= | |
| | \| | *primitives* |
| | \| | *f_typing* |
| | \| | *typing* |
| | \| | *translation* |

## 2.1 Functional simple type system FS

**Formulas equality** $\boxed{\tau = \tau'}$

$$\frac{}{\tau = \tau} \quad \text{(FORM_EQ_REFL)}$$

**Terms equality** $\boxed{t = t'}$

$$\frac{}{t = t} \quad \text{(TERM_EQ_REFL)}$$

**Lookup** $\boxed{x : \tau \in \Sigma}$

$$\frac{}{x : \tau \in \Sigma, x : \tau} \quad \text{(F_LOOKUP_I)}$$

$$\frac{x \neq y \qquad x : \tau \in \Sigma}{x : \tau \in \Sigma, y : \tau'} \quad \text{(F_LOOKUP_II)}$$

**Type check** $\boxed{\Sigma \vdash t : \tau}$

$$\frac{x : \tau \in \Sigma}{\Sigma \vdash x : \tau} \quad \text{(TC_VAR)}$$

$$\frac{}{\Sigma \vdash 0 : \mathbf{nat}} \quad \text{(TC_ZERO)}$$

$$\frac{\Sigma \vdash t : \mathbf{nat}}{\Sigma \vdash \mathbf{succ}\,(t) : \mathbf{nat}} \quad \text{(TC_SUCC)}$$

$$\frac{\Sigma \vdash t : \mathbf{nat}}{\Sigma \vdash \mathbf{pred}\,(t) : \mathbf{nat}} \quad \text{(TC_PRED)}$$

$$\frac{\Sigma, x : \tau \vdash t : \tau'}{\Sigma \vdash \mathbf{fn}\, x : \tau \Rightarrow t : \tau \to \tau'} \quad \text{(TC_LAM)}$$

$$\frac{\Sigma \vdash t_1 : \tau \to \tau' \qquad \Sigma \vdash t_2 : \tau}{\Sigma \vdash t_1\, t_2 : \tau'} \quad \text{(TC_APP)}$$

$$\frac{\Sigma \vdash t_1 : \mathbf{nat} \qquad \Sigma \vdash t_2 : \tau \qquad \Sigma \vdash t_3 : \mathbf{nat} \to (\tau \to \tau)}{\Sigma \vdash \mathbf{rec}\,(t_1, t_2, t_3) : \tau} \quad \text{(TC_REC)}$$

$$\frac{\Sigma \vdash (\vec{t}) : (\vec{\tau})}{\Sigma \vdash \langle \vec{t} \rangle : \langle \vec{\tau} \rangle} \quad \text{(TC_TUPLE)}$$

$$\frac{\Sigma \vdash t_1 : \tau \qquad \Sigma, y : \tau \vdash t_2 : \tau'}{\Sigma \vdash \mathbf{let}\, y = t_1\, \mathbf{in}\, t_2 : \tau'} \quad \text{(TC_LET)}$$

$$\frac{\Sigma \vdash t_1 : \tau \qquad \Sigma, \langle \vec{x} \rangle : \tau \vdash t_2 : \tau'}{\Sigma \vdash \mathbf{let}\, \langle \vec{x} \rangle = t_1\, \mathbf{in}\, t_2 : \tau'} \quad \text{(TC_MATCH)}$$

**Append** $\boxed{\Sigma, \vec{x} : \vec{\tau} = \Sigma'}$

$$\frac{}{\Sigma, () : () = \Sigma} \quad \text{(APP_I)}$$

$$\frac{\Sigma, \vec{x} : \vec{\tau} = \Sigma'}{\Sigma, (\vec{x}, x) : (\vec{\tau}, \tau) = \Sigma', x : \tau} \quad \text{(APP_II)}$$

**Type check term in extended environment** $\boxed{\Sigma, \langle \vec{x} \rangle : \tau \vdash t : \tau'}$

$$\frac{\Sigma, \vec{x} : \vec{\tau} = \Sigma' \qquad \Sigma' \vdash t : \tau'}{\Sigma, \langle \vec{x} \rangle : \langle \vec{\tau} \rangle \vdash t : \tau'} \qquad \text{(TCTE_PRODUCT)}$$

**Type check terms** $\boxed{\Sigma \vdash (\vec{t}) : (\vec{\tau})}$

$$\frac{}{\Sigma \vdash () : ()} \qquad \text{(TCTS_EMPTY)}$$

$$\frac{\Sigma \vdash t : \tau \qquad \Sigma \vdash (\vec{t}) : (\vec{\tau})}{\Sigma \vdash (\vec{t}, t) : (\vec{\tau}, \tau)} \qquad \text{(TCTS_CONS)}$$

## 2.2 Imperative simple type system IS

**Propositions equality** $\boxed{\tau = \tau'}$

$$\frac{}{\tau = \tau} \quad \text{(PROP_EQ_ID)}$$

**Lookup ident** $\boxed{x : \tau \in \Gamma}$

$$\frac{}{x : \tau \in \Gamma, x : \tau} \quad \text{(LOOKUP_I)}$$

$$\frac{x \neq x' \qquad x : \tau \in \Gamma}{x : \tau \in \Gamma, x' : \tau'} \quad \text{(LOOKUP_II)}$$

**Lookup idents** $\boxed{\vec{x} : \vec{\tau} \subset \Gamma}$

$$\frac{}{() : () \subset \Gamma} \quad \text{(LOOKUP_IDENTS_I)}$$

$$\frac{x : \tau \in \Gamma \qquad \vec{x} : \vec{\tau} \subset \Gamma}{\vec{x}, x : \vec{\tau}, \tau \subset \Gamma} \quad \text{(LOOKUP_IDENTS_II)}$$

**Update** $\boxed{\Omega[x : \tau] = \Omega'}$

$$\frac{}{(\Omega, x : \tau')[x : \tau] = (\Omega, x : \tau)} \quad \text{(UPDATE_I)}$$

$$\frac{x \neq x' \qquad \Omega[x : \tau] = \Omega'}{(\Omega, x' : \tau')[x : \tau] = (\Omega', x' : \tau')} \quad \text{(UPDATE_II)}$$

**Multi-update** $\boxed{\Omega[\![\vec{x} : \vec{\tau}]\!] = \Omega'}$

$$\frac{}{\Omega[\![() : ()]\!] = \Omega} \quad \text{(MULTI_UPDATE_I)}$$

$$\frac{\Omega[\![\vec{x} : \vec{\tau}]\!] = \Omega' \qquad \Omega'[x : \tau] = \Omega''}{\Omega[\![\vec{x}, x : \vec{\tau}, \tau]\!] = \Omega''} \quad \text{(MULTI_UPDATE_II)}$$

**Append** $\boxed{\Gamma, \gamma = \Gamma'}$

$$\frac{}{\Gamma, () = \Gamma} \quad \text{(APPEND_I)}$$

$$\frac{\Gamma, \gamma = \Gamma'}{\Gamma, (\gamma, x : \tau) = \Gamma', x : \tau} \quad \text{(APPEND_II)}$$

**Subset** $\boxed{\omega \subset \Omega}$

$$\frac{}{() \subset \Omega} \quad \text{(TC_SUBSET_I)}$$

$$\frac{\omega \subset \Omega \qquad x : \tau \in \Omega}{(\omega, x : \tau) \subset \Omega} \quad \text{(TC_SUBSET_II)}$$

**Restriction** $\boxed{\Omega_{|\vec{x}} = \omega}$

$$\frac{}{\Omega_{|()} = ()} \quad \text{(TC_RESTRICT_I)}$$

$$\frac{\Omega_{|\vec{x}} = \omega \qquad y : \tau \in \Omega}{\Omega_{|\vec{x}, y} = (\omega, y : \tau)} \quad \text{(TC_RESTRICT_II)}$$

**Split** $\boxed{\Omega = \vec{x} : \vec{\tau}}$

$$\frac{}{() = () : ()} \quad \text{(TC_SPLIT_I)}$$

$$\frac{\Omega = \vec{x} : \vec{\tau}}{(\Omega, x : \tau) = (\vec{x}, x) : (\vec{\tau}, \tau)} \quad \text{(TC_SPLIT_II)}$$

**Init** $\boxed{\vec{x} : \vec{\top} = \omega}$

$$\frac{}{() : \vec{\top} = ()} \quad \text{(TC_INIT_I)}$$

$$\frac{\vec{x} : \vec{\top} = \omega}{(\vec{x}, y) : \vec{\top} = (\omega, y : \top)} \quad \text{(TC_INIT_II)}$$

**Typecheck expression** $\boxed{\Gamma; \Omega \vdash e : \tau}$

$$\frac{x : \tau \in \Gamma}{\Gamma; \Omega \vdash x : \tau} \quad \text{(T_ENV_I)}$$

$$\frac{x : \tau \in \Omega}{\Gamma; \Omega \vdash x : \tau} \quad \text{(T_ENV_II)}$$

$$\frac{}{\Gamma; \Omega \vdash \star : \top} \quad \text{(T_UNIT)}$$

$$\frac{}{\Gamma; \Omega \vdash q : \mathbf{nat}} \quad \text{(T_NUM)}$$

$$\frac{\gamma = \vec{y} : \vec{\sigma} \quad \omega = \vec{z} : \vec{\tau} \quad \vec{z} : \vec{\top} = \omega' \quad \Gamma, \gamma = \Gamma' \quad \Gamma'; \omega' \vdash s \triangleright \omega}{\Gamma; \Omega \vdash \mathbf{proc}\,[\gamma]\,\mathbf{out}\,[\omega]\{s\} : \mathbf{proc}\,([\vec{\sigma}]\,\mathbf{out}\,[\vec{\tau}])} \quad \text{(T_PROC)}$$

**Typecheck expressions** $\boxed{\Gamma; \Omega \vdash (\vec{e}) : (\vec{\tau})}$

$$\frac{}{\Gamma; \Omega \vdash () : ()} \quad \text{(T_EXPS_I)}$$

$$\frac{\Gamma; \Omega \vdash (\vec{e}) : (\vec{\tau}) \quad \Gamma; \Omega \vdash e : \tau}{\Gamma; \Omega \vdash (\vec{e}, e) : (\vec{\tau}, \tau)} \quad \text{(T_EXPS_II)}$$

**Typecheck sequence** $\boxed{\Gamma; \Omega \vdash s \triangleright \Omega'}$

$$\frac{}{\Gamma; \Omega \vdash \varepsilon \triangleright \Omega} \quad \text{(T_EMPTY)}$$

$$\frac{\Gamma; \Omega \vdash e : \tau \quad \Gamma, y : \tau; \Omega \vdash s \triangleright \Omega'}{\Gamma; \Omega \vdash \mathbf{cst}\, y = e; s \triangleright \Omega'} \quad \text{(T_CST)}$$

$$\frac{\Gamma; \Omega \vdash e : \tau \quad \Gamma; \Omega, y : \tau \vdash s \triangleright \Omega', y : \tau'}{\Gamma; \Omega \vdash \mathbf{var}\, y := e; s \triangleright \Omega'} \quad \text{(T_VAR)}$$

$$\frac{\omega \subset \Omega \quad \omega = \vec{x} : \vec{\sigma} \quad \Gamma; \omega \vdash s \triangleright \omega' \quad \omega' = \vec{x} : \vec{\tau} \quad \Omega[\![\vec{x} : \vec{\tau}]\!] = \Omega' \quad \Gamma; \Omega' \vdash s' \triangleright \Omega''}{\Gamma; \Omega \vdash \{s\}_{\omega}; s' \triangleright \Omega''} \quad \text{(T_BLOCK)}$$

$$\frac{y : \mathbf{nat} \in \Omega \quad \Gamma; \Omega \vdash s \triangleright \Omega'}{\Gamma; \Omega \vdash \mathbf{inc}\,(y); s \triangleright \Omega'} \quad \text{(T_INC)}$$

$$\frac{y : \mathbf{nat} \in \Omega \qquad \Gamma; \Omega \vdash s \triangleright \Omega'}{\Gamma; \Omega \vdash \mathbf{dec}\,(y); s \triangleright \Omega'} \tag{T_DEC}$$

$$\frac{y : \tau \in \Omega \qquad \Gamma; \Omega \vdash e : \tau' \qquad \Omega[y : \tau'] = \Omega' \qquad \Gamma; \Omega' \vdash s \triangleright \Omega''}{\Gamma; \Omega \vdash y := e; s \triangleright \Omega''} \tag{T_ASSIGN}$$

$$\frac{\omega \subset \Omega \qquad \Gamma; \Omega \vdash e : \mathbf{nat} \qquad \Gamma, y : \mathbf{nat}; \omega \vdash s \triangleright \omega \qquad \Gamma; \Omega \vdash s' \triangleright \Omega'}{\Gamma; \Omega \vdash \mathbf{for}\, y := 0\, \mathbf{until}\, e\, \{s\}_\omega; s' \triangleright \Omega'} \tag{T_FOR}$$

$$\frac{\Gamma; \Omega \vdash p : \mathbf{proc}\,([\vec{\sigma}]\, \mathbf{out}\, [\vec{\tau}]) \qquad \Gamma; \Omega \vdash (\vec{e}) : (\vec{\sigma}) \qquad \Omega[\![\vec{z} : \vec{\tau}]\!] = \Omega' \qquad \Gamma; \Omega' \vdash s \triangleright \Omega''}{\Gamma; \Omega \vdash p(\vec{e}; \vec{z}); s \triangleright \Omega''} \tag{T_CALL}$$

## 2.3 Translation from IS to FS

**Types translation** $\boxed{(\tau)^\star = \tau}$

$$\frac{}{(\mathbf{nat})^\star = \mathbf{nat}} \quad (\text{TR_TYPE_1})$$

$$\frac{}{(\top)^\star = \top} \quad (\text{TR_TYPE_2})$$

$$\frac{(\vec{\tau})^\star = (\vec{\tau}) \qquad (\vec{\tau}')^\star = (\vec{\tau}')}{(\mathbf{proc}\,([\vec{\tau}]\,\mathbf{out}\,[\vec{\tau}']))^\star = \langle\vec{\tau}\rangle \to \langle\vec{\tau}'\rangle} \quad (\text{TR_TYPE_3})$$

**Types translation** $\boxed{(\vec{\tau})^\star = (\vec{\tau})}$

$$\frac{}{()^\star = ()} \quad (\text{TR_TYPES_1})$$

$$\frac{(\vec{\tau})^\star = (\vec{\tau}) \qquad (\tau)^\star = \tau}{(\vec{\tau}, \tau)^\star = (\vec{\tau}, \tau)} \quad (\text{TR_TYPES_2})$$

**Sequence translation** $\boxed{(\vec{x})^\star = \vec{t}}$

$$\frac{}{()^\star = ()} \quad (\text{TR_IDENTS_1})$$

$$\frac{(\vec{x})^\star = \vec{t}}{(\vec{x}, x)^\star = (\vec{t}, x)} \quad (\text{TR_IDENTS_2})$$

**Number translation** $\boxed{q^\star = t}$

$$\frac{}{0^\star = 0} \quad (\text{TR_NUM_1})$$

$$\frac{q^\star = t}{\mathbf{succ}\,(q)^\star = \mathbf{succ}\,(t)} \quad (\text{TR_NUM_2})$$

**Expression translation** $\boxed{(e)^\star = t}$

$$\frac{q^\star = t}{(q)^\star = t} \quad (\text{TR_EXP_1})$$

$$\frac{}{(x)^\star = x} \quad (\text{TR_EXP_2})$$

$$\frac{}{(\star)^\star = \langle\rangle} \quad (\text{TR_EXP_3})$$

$$\frac{\omega = \vec{z} : \vec{\tau} \qquad (s)^\star_{\vec{z}} = t \qquad \gamma = \vec{x} : \vec{\sigma} \qquad (\vec{\sigma})^\star = (\vec{\tau})}{(\mathbf{proc}\,[\gamma]\,\mathbf{out}\,[\omega]\{s\})^\star = \mathbf{fn}\,(\vec{x} : \vec{\tau}) \Rightarrow t} \quad (\text{TR_EXP_4})$$

**Expressions translation** $\boxed{(\vec{e})^\star = \vec{t}}$

$$\frac{}{()^\star =} \quad (\text{TR_EXPS_I})$$

$$\frac{(\vec{e})^\star = \vec{t} \qquad (e)^\star = t}{(\vec{e}, e)^\star = \vec{t}, t} \quad (\text{TR_EXPS_II})$$

**Sequence translation** $\boxed{(s)^\star_{\vec{x}} = t}$

$$\frac{(\vec{x})^\star = \vec{t}}{()^\star_{\vec{x}} = \langle\vec{t}\rangle} \qquad (\text{TR_SEQ_1})$$

$$\frac{(e)^\star = t \qquad (s)^\star_{\vec{x}} = t'}{(\mathbf{var}\, x := e; s)^\star_{\vec{x}} = \mathbf{let}\, x = t\, \mathbf{in}\, t'} \qquad (\text{TR_SEQ_2})$$

$$\frac{(e)^\star = t \qquad (s)^\star_{\vec{x}} = t'}{(\mathbf{cst}\, x = e; s)^\star_{\vec{x}} = \mathbf{let}\, x = t\, \mathbf{in}\, t'} \qquad (\text{TR_SEQ_3})$$

$$\frac{(e)^\star = t \qquad (s)^\star_{\vec{x}} = t'}{(x := e; s)^\star_{\vec{x}} = \mathbf{let}\, x = t\, \mathbf{in}\, t'} \qquad (\text{TR_SEQ_4})$$

$$\frac{(s)^\star_{\vec{x}} = t}{(\mathbf{inc}\,(x); s)^\star_{\vec{x}} = \mathbf{let}\, x = \mathbf{succ}\,(x)\, \mathbf{in}\, t} \qquad (\text{TR_SEQ_5})$$

$$\frac{(s)^\star_{\vec{x}} = t}{(\mathbf{dec}\,(x); s)^\star_{\vec{x}} = \mathbf{let}\, x = \mathbf{pred}\,(x)\, \mathbf{in}\, t} \qquad (\text{TR_SEQ_6})$$

$$\frac{(e)^\star = t \qquad (\vec{e})^\star = \vec{u} \qquad (s)^\star_{\vec{x}} = t'}{(e(\vec{e}; \vec{z}); s)^\star_{\vec{x}} = \mathbf{let}\, \langle\vec{z}\rangle = t\, \langle\vec{u}\rangle\, \mathbf{in}\, t'} \qquad (\text{TR_SEQ_7})$$

$$\frac{\omega = \vec{z} : \vec{\sigma} \qquad (s_1)^\star_{\vec{z}} = t_1 \qquad (s_2)^\star_{\vec{x}} = t_2}{(\{s_1\}_\omega; s_2)^\star_{\vec{x}} = \mathbf{let}\, \langle\vec{z}\rangle = t_1\, \mathbf{in}\, t_2} \qquad (\text{TR_SEQ_8})$$

$$\frac{\omega = \vec{z} : \vec{\sigma} \qquad (\vec{z})^\star = \vec{u} \qquad (\vec{\sigma})^\star = (\vec{\tau}) \qquad (e)^\star = t_0 \qquad (s_1)^\star_{\vec{z}} = t_1 \qquad (s_2)^\star_{\vec{x}} = t_2}{(\mathbf{for}\, y := 0\, \mathbf{until}\, e\, \{s_1\}_\omega; s_2)^\star_{\vec{x}} = \mathbf{let}\, \langle\vec{z}\rangle = \mathbf{rec}\,(t_0, \langle\vec{u}\rangle, \mathbf{fn}\, y : \mathbf{nat} \Rightarrow \mathbf{fn}\,(\vec{z} : \vec{\tau}) \Rightarrow t_1)\, \mathbf{in}\, t_2} \qquad (\text{TR_SEQ_9})$$

# 3 Grammars and judgments for FD and ID

| | | | | |
|---|---|---|---|---|
| *ident*, $x$, $y$, $z$ | ::= | | | variable: |
| | \| | *variable* | | |
| | | | | |
| *idents*, $\vec{x}$, $\vec{y}$, $\vec{z}$ | ::= | | | variables: |
| | \| | | | |
| | \| | $\vec{x}, x$ | | |
| | \| | $x$ | S | |
| | \| | $(\vec{x})$ | S | |
| | | | | |
| *fenv*, $\Sigma$ | ::= | | | Environments: |
| | \| | $\{\}$ | | empty environment |
| | \| | $\Sigma, x : \varphi$ | | ident declaration |
| | | | | |
| *terms*, $\vec{t}$, $\vec{u}$ | ::= | | | variables: |
| | \| | | S | |
| | \| | $\vec{t}, t$ | S | |
| | \| | $t_1, t_2$ | S | |
| | \| | $(\vec{t})$ | S | |
| | | | | |
| *term*, $t$, $u$ | ::= | | | term: |
| | \| | $x$ | | var |
| | \| | $0$ | | zero |
| | \| | $1$ | S | |
| | \| | $2$ | S | |
| | \| | $3$ | S | |
| | \| | $4$ | S | |
| | \| | $5$ | S | |
| | \| | $t_1\ t_2$ | | application |
| | \| | $\mathbf{fn}\, x : \varphi \Rightarrow t$ | | abstraction |
| | \| | $\mathbf{fn}\, (\vec{x} : \vec{\varphi}) \Rightarrow t$ | | multi-abstraction |
| | \| | $t[i]$ | S | meta-application |
| | \| | $\lambda n.t$ | S | generalization |
| | \| | $?n.t$ | S | any |
| | \| | $t\{i\}$ | | instance |
| | \| | $\mathbf{succ}\,(t)$ | | successor |
| | \| | $\mathbf{pred}\,(t)$ | | predecessor |
| | \| | $\mathbf{rec}(t_1, t_2, t_3)$ | | recursor |
| | \| | $\mathbf{let}\, x = t_1\, \mathbf{in}\, t_2$ | | let |
| | \| | $i_1 = i_2$ | | axiom |
| | \| | $t :> \varphi[t']$ | S | subst |
| | \| | $\langle i, t : \varphi \rangle$ | | witness |
| | \| | $\langle \vec{t} \rangle$ | | tuple |
| | \| | $\langle t \rangle$ | | degenerated tuple |
| | \| | $\mathbf{let}\, \langle \vec{x} \rangle = t_1\, \mathbf{in}\, t_2$ | | match |
| | \| | $\mathbf{throw}_{\varphi}\ t_1\ t_2$ | | throw |
| | \| | $\mathbf{callcc}\, t$ | | callcc |
| | \| | $(t)$ | S | |
| | | | | |
| *form*, $\varphi$ | ::= | | | formula: |
| | \| | $x$ | | var |
| | \| | $\top$ | | true |
| | \| | $\bot$ | | false |

| | | | | |
|---|---|---|---|---|
| | \| | $\mathbf{nat}(i)$ | | nat |
| | \| | $i = i'$ | | equals |
| | \| | $\varphi \to \varphi'$ | | imply |
| | \| | $\neg\varphi$ | | not |
| | \| | $\forall n\ \varphi$ | S | forall |
| | \| | $\exists n\ \varphi$ | S | exists |
| | \| | $\varphi[i]$ | S | meta-application |
| | \| | $\{n/\varphi\}$ | S | meta-abstraction |
| | \| | $\varphi[x = i]$ | S | meta-substitution |
| | \| | $\langle\vec{\varphi}\rangle$ | | tuple |
| | \| | $(\varphi)$ | S | |
| *forms*, $\vec{\varphi}$ | ::= | | | formulas: |
| | \| | | S | |
| | \| | $\varphi$ | S | |
| | \| | $\vec{\varphi}, \varphi$ | S | |
| | \| | $\vec{\varphi}[i]$ | S | meta-application |
| | \| | $(\vec{\varphi})$ | S | |
| *absterm*, $t$ | ::= | | | Parametrized term: |
| | \| | $n \mapsto t$ | S | |
| *absforms*, $\vec{\varphi}$ | ::= | | | Parametrized formulas: |
| | \| | $n \mapsto \vec{\varphi}$ | S | |
| *ind*, $i$ | ::= | | | individuals: |
| | \| | 0 | | zero |
| | \| | 1 | S | |
| | \| | 2 | S | |
| | \| | 3 | S | |
| | \| | 4 | S | |
| | \| | 5 | S | |
| | \| | $succ(i)$ | | successor |
| | \| | $pred(i)$ | | predecessor |
| | \| | $add(i_1, i_2)$ | | addition |
| | \| | $sub(i_1, i_2)$ | | subtraction |
| | \| | $mult(i_1, i_2)$ | | multiplication |
| | \| | $F_{32}(i)$ | | F32 |
| | \| | $n$ | S | variable |
| *env*, $\Gamma$, $\Omega$, $\gamma$, $\omega$ | ::= | | | Environment: |
| | \| | | | empty environment |
| | \| | $\Gamma, x : \psi$ | | ident declaration |
| | \| | $x : \psi$ | S | ident declaration |
| | \| | $\Gamma[i]$ | S | meta-application |
| | \| | $\{n/\Gamma\}$ | S | meta-abstraction |
| | \| | $\Gamma[n = i]$ | S | meta-substitution |
| | \| | $(\Gamma)$ | S | |
| *absenv*, $\theta$ | ::= | | | Parametrized existentially quantified environments: |
| | \| | $n \mapsto \Gamma$ | S | |
| *qenv*, $\Theta$, $\theta$ | ::= | | | Existentially quantified environments: |
| | \| | $[\Omega]$ | | simple |
| | \| | $\exists n\ \Theta$ | S | binder |
| | \| | $\Theta[i]$ | S | meta-application |
| | \| | $(\Theta)$ | S | |

| | | | | |
|---|---|---|---|---|
| *absqenv*, $\theta$ | ::= | | | Parametrized existentially quantified environments: |
| | \| | $\{x/\Theta\}$ | S | |
| *command*, $c$ | ::= | | | Command: |
| | \| | $\{s\}_\theta$ | | block |
| | \| | $\textbf{for}\, y : \textbf{nat}\,(n) := 0\,\textbf{until}\, e\{s\}[\omega]$ | S | for |
| | \| | $\textbf{for}\ y : \textbf{nat}(n) := 0\ \textbf{until}\ e\ \{s\}_\omega$ | S | for |
| | \| | $y := e$ | | assign |
| | \| | $\textbf{inc}\,(y)$ | | inc |
| | \| | $\textbf{dec}\,(y)$ | | dec |
| | \| | $e(\vec{e}; \vec{y})$ | | call |
| | \| | $\textbf{jump}\ (e, \vec{e})_\theta$ | | jump |
| | \| | $y : \{s\}_\theta$ | | label |
| *sequence*, $s$ | ::= | | | Sequence: |
| | \| | | S | implicit empty sequence |
| | \| | $\varepsilon$ | S | explicit empty sequence |
| | \| | $c; s$ | S | command |
| | \| | $\textbf{cst}\, y = e; s$ | S | constant |
| | \| | $\textbf{var}\, y := e; s$ | S | variable |
| | \| | $\textbf{var}\, y; s$ | S | variable |
| | \| | $s[i]$ | S | meta-application |
| | \| | $?n.s$ | S | abstraction |
| | \| | $[i \in \theta]s$ | S | witness |
| | \| | $s :> \theta[e]$ | S | subst |
| | \| | $(s)$ | S | |
| *body*, $b$ | ::= | | | Parametrized sequence: |
| | \| | $n \mapsto s$ | S | meta-abstraction |
| *number*, $q$ | ::= | | | Number: |
| | \| | $0$ | | zero |
| | \| | $1$ | S | |
| | \| | $2$ | S | |
| | \| | $3$ | S | |
| | \| | $4$ | S | |
| | \| | $5$ | S | |
| | \| | $\mathbf{s}(q)$ | | successor |
| *expression*, $e$ | ::= | | | Expression: |
| | \| | $x$ | | variable |
| | \| | $\star$ | | star |
| | \| | $q$ | | number |
| | \| | $e[i]$ | S | meta-application |
| | \| | $e\{i\}$ | | procedure instance |
| | \| | $e <: \vec{\phi}\{i\}$ | | continuation instance |
| | \| | $e :> \vec{\psi}[e']$ | S | subst |
| | \| | $i_1 = i_2$ | | axiom |
| | \| | $\textbf{proc}\, h$ | | procedure |
| *header*, $h$ | ::= | | | Header: |
| | \| | $[\gamma]\,\textbf{out}\,\theta\{s\}$ | | parameters |
| | \| | $h[i]$ | S | meta-application |
| | \| | $\forall n\ h$ | S | generalization |
| *expressions*, $\vec{e}$ | ::= | | | Expressions: |
| | \| | | S | |
| | \| | $\vec{e}, e$ | S | |

| | | | | |
|---|---|---|---|---|
| | \| | $e$ | S | |
| | \| | $(\vec{e})$ | S | |
| *prop*, $\psi$, $\rho$ | ::= | | | Dependent type: |
| | \| | $x$ | | var |
| | \| | $i_1 = i_2$ | | equality |
| | \| | $\top$ | | true |
| | \| | $\top$ | S | true |
| | \| | $\bot$ | | false |
| | \| | $\bot$ | S | false |
| | \| | $\mathbf{nat}\,(i)$ | | nat |
| | \| | $\mathbf{proc}\,\rho$ | | proc |
| | \| | $\sim\vec{\psi}$ | S | |
| | \| | $\psi[i]$ | S | meta-application |
| | \| | $(\psi)$ | S | |
| *absprop*, $\vec{\psi}$ | ::= | | | Parametrized dependent type: |
| | \| | $\{n/\psi\}$ | S | |
| *props*, $\vec{\psi}$, $\vec{\rho}$ | ::= | | | Dependent types: |
| | \| | | S | |
| | \| | $\vec{\psi}, \psi$ | S | |
| | \| | $\psi$ | S | |
| | \| | $\vec{\psi}[i]$ | S | meta-application |
| | \| | $(\vec{\psi})$ | S | |
| *absprops*, $\vec{\psi}$ | ::= | | | Parametrized dependent types: |
| | \| | $n \mapsto \vec{\psi}$ | S | |
| *output*, $\phi$ | ::= | | | Existentially quantified dependent type: |
| | \| | $[\vec{\psi}]$ | | dependent types |
| | \| | $\exists n\ \phi$ | S | existential quantification |
| | \| | $\phi[i]$ | S | meta-application |
| | \| | $(\phi)$ | S | |
| *absoutput*, $\vec{\phi}$ | ::= | | | Parametrized existentially quantified dependent type: |
| | \| | $\{n/\phi\}$ | S | |
| *prototype*, $\rho$ | ::= | | | Universally quantified prototype: |
| | \| | $([\vec{\psi}]\,\mathbf{out}\,\phi)$ | | in/out parameters |
| | \| | $\forall n\ \rho$ | S | universal quantification |
| | \| | $\rho[i]$ | S | meta-application |
| | \| | $\{n/\rho\}$ | S | meta-abstraction |
| *primitives* | ::= | | | |
| *axiomes* | ::= | | | |
| | \| | $\vdash i = i'$ | | Axioms |
| *f_typing* | ::= | | | |
| | \| | $\varphi = \varphi'$ | | Formulas equality |
| | \| | $x : \varphi \in \Sigma$ | | Lookup |
| | \| | $\Sigma \vdash t : \varphi$ | | Type check term |
| | \| | $\Sigma \vdash (\vec{t}) : (\vec{\varphi})$ | | Type check terms |
| | \| | $\Sigma, \vec{x} : \vec{\varphi} = \Sigma'$ | | Append environments |
| | \| | $\Sigma, \langle\vec{x}\rangle : \varphi \vdash t : \varphi'$ | | Type check term in extended environment |

| | | | |
|---|---|---|---|
| *typing* | ::= | | |
| | \| | $\psi = \psi'$ | Formula equality |
| | \| | $\gamma = \gamma'$ | Environment equality |
| | \| | $x : \psi \in \Gamma$ | Lookup ident |
| | \| | $x \notin \Gamma$ | Not in environment |
| | \| | $y \notin \Theta$ | Not in quantified environment |
| | \| | $y \in \Theta$ | Lookup ident |
| | \| | $\vec{x} : \vec{\psi} \subset \Gamma$ | Lookup idents |
| | \| | $\Omega[x : \psi] = \Omega'$ | Update |
| | \| | $\Omega[\![\vec{x} : \vec{\psi}]\!] = \Omega'$ | Multi-update |
| | \| | $\Gamma; \Omega[x : \psi] \vdash s \triangleright \Theta$ | Type check with updated environment |
| | \| | $\Gamma; \Omega[\![\vec{x} : \vec{\psi}]\!] \vdash s \triangleright \Theta$ | Type check with updated environment |
| | \| | $\Gamma; \Omega[\![\omega]\!] \vdash s \triangleright \Theta$ | Type check with updated environment |
| | \| | $\Gamma, \gamma = \Gamma'$ | Append |
| | \| | $\omega \subset \Omega$ | Subset |
| | \| | $\Omega_{\|\vec{x}} = \omega$ | Restriction |
| | \| | $\Omega = \vec{x} : \vec{\psi}$ | Split |
| | \| | $\Theta = \vec{x} : \phi$ | Split quantified environment |
| | \| | $\Omega \Leftarrow \vec{x} : \vec{\psi}$ | Zip |
| | \| | $\Theta \Leftarrow \vec{x} : \phi$ | Zip quantified environment |
| | \| | $\vec{x} : \psi = \omega$ | Init |
| | \| | $\Gamma; \Omega \vdash e : \psi$ | Typecheck expression |
| | \| | $\Gamma; \Omega \vdash (\vec{e}) : (\vec{\psi})$ | Typecheck expressions |
| | \| | $\sim \phi = \psi$ | Defined negation |
| | \| | $\Gamma; \Omega \vdash s \triangleright \Theta$ | Typecheck sequence |
| | \| | $\Gamma; \Omega[\![\Theta]\!] \vdash s \triangleright \Theta'$ | Typecheck sequence with updated environment |
| *translation* | ::= | | |
| | \| | $\psi^\star = \varphi$ | Type translation |
| | \| | $(\vec{\psi})^\star = (\vec{\varphi})$ | Types translation |
| | \| | $(\gamma)^\star = (\vec{x}) : (\vec{\varphi})$ | Environment translation |
| | \| | $(\theta)^\star = \vec{z} : \vec{\varphi}$ | Parametrized environment translation |
| | \| | $(\vec{\psi})^\star = \varphi$ | Parametrized type translation |
| | \| | $(\vec{\psi})^\star = (\vec{\varphi})$ | Parametrized types translation |
| | \| | $(\phi)^\star = \varphi$ | Quantified types translation |
| | \| | $(\theta)^\star = \langle\vec{x}\rangle : \varphi$ | Quantified types translation |
| | \| | $(\rho)^\star = \varphi$ | Prototype translation |
| | \| | $(\vec{x})^\star = \vec{t}$ | Idents translation |
| | \| | $q^\star = t$ | Number translation |
| | \| | $(h)^\star = t$ | Header translation |
| | \| | $(e)^\star = t$ | Expression translation |
| | \| | $(\vec{e})^\star = (\vec{t})$ | Expressions translation |
| | \| | $(s)^\star_{\vec{x}} = t$ | Sequence translation |
| | \| | $(b)^\star_{\vec{x}} = t$ | Loop body translation |
| *judgement* | ::= | | |
| | \| | *primitives* | |
| | \| | *axiomes* | |
| | \| | *f_typing* | |
| | \| | *typing* | |
| | \| | *translation* | |

# Axioms

**Axioms** $\boxed{\vdash i = i'}$

$$\frac{}{\vdash i = i} \quad \text{(AX_REFL)}$$

$$\frac{}{\vdash \mathit{pred}(0) = 0} \quad \text{(AX_PRED_0)}$$

$$\overline{\vdash pred(succ(i)) = i} \quad (\textsc{ax_pred_s})$$

$$\overline{\vdash add(0, i') = i'} \quad (\textsc{ax_add_0})$$

$$\overline{\vdash add(succ(i), i') = succ(add(i, i'))} \quad (\textsc{ax_add_s})$$

$$\overline{\vdash add(i', 0) = i'} \quad (\textsc{ax_add2_0})$$

$$\overline{\vdash add(i', succ(i)) = succ(add(i', i))} \quad (\textsc{ax_add2_s})$$

$$\overline{\vdash mult(0, i') = i'} \quad (\textsc{ax_mult_0})$$

$$\overline{\vdash mult(succ(i), i') = add(mult(i, i'), i')} \quad (\textsc{ax_mult_s})$$

$$\overline{\vdash F_{32}(0) = 3} \quad (\textsc{ax_F32_0})$$

$$\overline{\vdash F_{32}(succ(i)) = 2} \quad (\textsc{ax_F32_s})$$

## 3.1 Functional dependent type system FD

**Formulas equality** $\boxed{\varphi = \varphi'}$

$$\frac{}{\varphi = \varphi} \quad \text{(FORM_EQ_I)}$$

**Lookup** $\boxed{x : \varphi \in \Sigma}$

$$\frac{}{x : \varphi \in \Sigma, x : \varphi} \quad \text{(F_LOOKUP_I)}$$

$$\frac{x \neq y \qquad x : \varphi \in \Sigma}{x : \varphi \in \Sigma, y : \varphi'} \quad \text{(F_LOOKUP_II)}$$

**Type check term** $\boxed{\Sigma \vdash t : \varphi}$

$$\frac{x : \varphi \in \Sigma}{\Sigma \vdash x : \varphi} \quad \text{(TC_VAR)}$$

$$\frac{}{\Sigma \vdash 0 : \mathbf{nat}(0)} \quad \text{(TC_ZERO)}$$

$$\frac{\Sigma \vdash t : \mathbf{nat}(i)}{\Sigma \vdash \mathbf{succ}\,(t) : \mathbf{nat}(succ(i))} \quad \text{(TC_SUCC)}$$

$$\frac{\Sigma, x : \varphi \vdash t : \varphi'}{\Sigma \vdash \mathbf{fn}\, x : \varphi \Rightarrow t : \varphi \to \varphi'} \quad \text{(TC_LAM)}$$

$$\frac{\Sigma \vdash t_1 : \varphi \to \varphi' \qquad \Sigma \vdash t_2 : \varphi}{\Sigma \vdash t_1\ t_2 : \varphi'} \quad \text{(TC_APP)}$$

$$\frac{\forall I \cdot \Sigma \vdash t[I] : \varphi[I]}{\Sigma \vdash \lambda n.t[n] : \forall n\ \varphi[n]} \quad \text{(TC_FORALL_I)}$$

$$\frac{\Sigma \vdash t : \forall n\ \varphi[n]}{\Sigma \vdash t\{i\} : \varphi[i]} \quad \text{(TC_FORALL_E)}$$

$$\frac{\Sigma \vdash (\vec{t}) : (\vec{\varphi})}{\Sigma \vdash \langle \vec{t} \rangle : \langle \vec{\varphi} \rangle} \quad \text{(TC_TUPLE)}$$

$$\frac{\Sigma \vdash t_1 : \varphi \qquad \Sigma, y : \varphi \vdash t_2 : \varphi'}{\Sigma \vdash \mathbf{let}\ y = t_1\ \mathbf{in}\ t_2 : \varphi'} \quad \text{(TC_LET)}$$

$$\frac{\Sigma \vdash t_1 : \varphi \qquad \Sigma, \langle \vec{x} \rangle : \varphi \vdash t_2 : \varphi'}{\Sigma \vdash \mathbf{let}\ \langle \vec{x} \rangle = t_1\ \mathbf{in}\ t_2 : \varphi'} \quad \text{(TC_MATCH)}$$

$$\frac{\Sigma \vdash t : \varphi[i]}{\Sigma \vdash \langle i, t : \exists n\ \varphi[n] \rangle : \exists n\ \varphi[n]} \quad \text{(TC_EXISTS_I)}$$

$$\frac{\Sigma \vdash t_1 : \mathbf{nat}(i) \qquad \Sigma \vdash t_2 : \varphi[0] \qquad \forall N \cdot \Sigma, y : \mathbf{nat}(N) \vdash t_3[N] : \varphi[N] \to \varphi[succ(N)]}{\Sigma \vdash \mathbf{rec}(t_1, t_2, \lambda n.\mathbf{fn}\, y : \mathbf{nat}(n) \Rightarrow t_3[n]) : \varphi[i]} \quad \text{(TC_REC)}$$

$$\frac{\vdash i_1 = i_2}{\Sigma \vdash i_1 = i_2 : i_1 = i_2} \quad \text{(TC_AX_I)}$$

$$\frac{\vdash i_1 = i_2}{\Sigma \vdash i_2 = i_1 : i_2 = i_1} \quad \text{(TC_AX_II)}$$

$$\frac{\Sigma \vdash t : \varphi[i_2] \qquad \Sigma \vdash t' : i_1 = i_2}{\Sigma \vdash t :> \exists n\ \varphi[n][t'] : \varphi[i_1]} \qquad \text{(TC_EQUAL_E)}$$

$$\frac{\Sigma \vdash t_1 : \neg\varphi \qquad \Sigma \vdash t_2 : \varphi}{\Sigma \vdash \mathbf{throw}_{\varphi'}\ t_1\ t_2 : \varphi'} \qquad \text{(TC_THROW)}$$

$$\frac{\Sigma \vdash t : \neg\varphi \to \varphi}{\Sigma \vdash \mathbf{callcc}\, t : \varphi} \qquad \text{(TC_CALLCC)}$$

**Type check terms** $\boxed{\Sigma \vdash (\vec{t}) : (\vec{\varphi})}$

$$\frac{}{\Sigma \vdash () : ()} \qquad \text{(TC_EMPTY)}$$

$$\frac{\Sigma \vdash t : \varphi \qquad \Sigma \vdash (\vec{t}) : (\vec{\varphi})}{\Sigma \vdash (\vec{t}, t) : (\vec{\varphi}, \varphi)} \qquad \text{(TC_CONS)}$$

**Append environments** $\boxed{\Sigma, \vec{x} : \vec{\varphi} = \Sigma'}$

$$\frac{}{\Sigma, () : () = \Sigma} \qquad \text{(APP_I)}$$

$$\frac{\Sigma, \vec{x} : \vec{\varphi} = \Sigma'}{\Sigma, (\vec{x}, x) : (\vec{\varphi}, \varphi) = \Sigma', x : \varphi} \qquad \text{(APP_II)}$$

**Type check term in extended environment** $\boxed{\Sigma, \langle\vec{x}\rangle : \varphi \vdash t : \varphi'}$

$$\frac{\Sigma, \vec{x} : \vec{\varphi} = \Sigma' \qquad \Sigma' \vdash t : \varphi'}{\Sigma, \langle\vec{x}\rangle : \langle\vec{\varphi}\rangle \vdash t : \varphi'} \qquad \text{(TC_PRODUCT)}$$

$$\frac{\forall I \cdot \Sigma, \langle\vec{x}\rangle : \varphi[I] \vdash t[I] : \varphi'}{\Sigma, \langle\vec{x}\rangle : \exists n\ \varphi[n] \vdash ?n.t[n] : \varphi'} \qquad \text{(TC_EXISTS)}$$

## 3.2 Imperative dependent type system ID

**Formula equality** $\boxed{\psi = \psi'}$

$$\frac{}{\psi = \psi} \quad \text{(PROP_EQ_ID)}$$

**Environment equality** $\boxed{\gamma = \gamma'}$

$$\frac{}{\gamma = \gamma} \quad \text{(ENV_EQ_ID)}$$

**Lookup ident** $\boxed{x : \psi \in \Gamma}$

$$\frac{}{x : \psi \in \Gamma, x : \psi} \quad \text{(LOOKUP_I)}$$

$$\frac{x \neq x' \qquad x : \psi \in \Gamma}{x : \psi \in \Gamma, x' : \psi'} \quad \text{(LOOKUP_II)}$$

**Not in environment** $\boxed{x \notin \Gamma}$

$$\frac{}{x \notin ()} \quad \text{(NOTIN_I)}$$

$$\frac{x \neq x' \qquad x \notin \Gamma}{x \notin \Gamma, x' : \psi'} \quad \text{(NOTIN_II)}$$

**Not in quantified environment** $\boxed{y \notin \Theta}$

$$\frac{y \notin \Gamma}{y \notin [\Gamma]} \quad \text{(NOTIN_QENVI)}$$

$$\frac{\forall I \cdot y \notin \Theta[I]}{y \notin \exists n\ \Theta[n]} \quad \text{(NOTIN_QENVII)}$$

**Lookup ident** $\boxed{y \in \Theta}$

$$\frac{y : \psi \in \Gamma}{y \in [\Gamma]} \quad \text{(BELONGS_I)}$$

$$\frac{\forall I \cdot y \in \Theta[I]}{y \in \exists n\ \Theta[n]} \quad \text{(BELONGS_II)}$$

**Lookup idents** $\boxed{\vec{x} : \vec{\psi} \subset \Gamma}$

$$\frac{}{() : () \subset \Gamma} \quad \text{(LOOKUP_IDENTS_I)}$$

$$\frac{x : \psi \in \Gamma \qquad \vec{x} : \vec{\psi} \subset \Gamma}{\vec{x}, x : \vec{\psi}, \psi \subset \Gamma} \quad \text{(LOOKUP_IDENTS_II)}$$

**Update** $\boxed{\Omega[x : \psi] = \Omega'}$

$$\frac{}{(\Omega, x : \psi')[x : \psi] = (\Omega, x : \psi)} \quad \text{(UPDATE_I)}$$

$$\frac{x \neq x' \qquad \Omega[x : \psi] = \Omega'}{(\Omega, x' : \psi')[x : \psi] = (\Omega', x' : \psi')} \quad \text{(UPDATE_II)}$$

**Multi-update** $\boxed{\Omega[\![\vec{x} : \vec{\psi}]\!] = \Omega'}$

$$\overline{\Omega[\![() : ()]\!] = \Omega} \quad \text{(MULTI_UPDATE_I)}$$

$$\frac{\Omega[\![\vec{x} : \vec{\psi}]\!] = \Omega' \qquad \Omega'[x : \psi] = \Omega''}{\Omega[\![\vec{x}, x : \vec{\psi}, \psi]\!] = \Omega''} \quad \text{(MULTI_UPDATE_II)}$$

**Type check with updated environment** $\boxed{\Gamma; \Omega[x : \psi] \vdash s \triangleright \Theta}$

$$\frac{\Omega[x : \psi] = \Omega' \qquad \Gamma; \Omega' \vdash s \triangleright \Theta}{\Gamma; \Omega[x : \psi] \vdash s \triangleright \Theta} \quad \text{(PRE_UPDATE_I)}$$

**Type check with updated environment** $\boxed{\Gamma; \Omega[\![\vec{x} : \vec{\psi}]\!] \vdash s \triangleright \Theta}$

$$\frac{\Omega[\![\vec{x} : \vec{\psi}]\!] = \Omega' \qquad \Gamma; \Omega' \vdash s \triangleright \Theta}{\Gamma; \Omega[\![\vec{x} : \vec{\psi}]\!] \vdash s \triangleright \Theta} \quad \text{(M_PRE_UPDATE_I)}$$

**Type check with updated environment** $\boxed{\Gamma; \Omega[\![\omega]\!] \vdash s \triangleright \Theta}$

$$\frac{\omega = \vec{x} : \vec{\psi} \qquad \Gamma; \Omega[\![\vec{x} : \vec{\psi}]\!] \vdash s \triangleright \Theta}{\Gamma; \Omega[\![\omega]\!] \vdash s \triangleright \Theta} \quad \text{(M_UPDATE_SHORT_I)}$$

**Append** $\boxed{\Gamma, \gamma = \Gamma'}$

$$\overline{\Gamma, () = \Gamma} \quad \text{(APPEND_I)}$$

$$\frac{\Gamma, \gamma = \Gamma'}{\Gamma, (\gamma, x : \psi) = \Gamma', x : \psi} \quad \text{(APPEND_II)}$$

**Subset** $\boxed{\omega \subset \Omega}$

$$\overline{() \subset \Omega} \quad \text{(TC_SUBSET_I)}$$

$$\frac{\omega \subset \Omega \qquad x : \psi \in \Omega}{(\omega, x : \psi) \subset \Omega} \quad \text{(TC_SUBSET_II)}$$

**Restriction** $\boxed{\Omega_{|\vec{x}} = \omega}$

$$\overline{\Omega_{|()} = ()} \quad \text{(TC_RESTRICT_I)}$$

$$\frac{\Omega_{|\vec{x}} = \omega \qquad y : \psi \in \Omega}{\Omega_{|\vec{x}, y} = (\omega, y : \psi)} \quad \text{(TC_RESTRICT_II)}$$

**Split** $\boxed{\Omega = \vec{x} : \vec{\psi}}$

$$\overline{() = () : ()} \quad \text{(TC_SPLIT_I)}$$

$$\frac{\Omega = \vec{x} : \vec{\psi}}{(\Omega, x : \psi) = (\vec{x}, x) : (\vec{\psi}, \psi)} \quad \text{(TC_SPLIT_II)}$$

**Split quantified environment** $\boxed{\Theta = \vec{x} : \phi}$

$$\frac{\Omega = \vec{x} : \vec{\psi}}{[\Omega] = \vec{x} : [\vec{\psi}]} \quad \text{(TC_QSPLIT_I)}$$

$$\frac{\forall N \cdot (\Theta[N] = \vec{x} : \phi[N])}{\exists n\ \Theta[n] = \vec{x} : \exists n\ \phi[n]} \quad \text{(TC_QSPLIT_II)}$$

**Zip** $\boxed{\Omega \Leftarrow \vec{x} : \vec{\psi}}$

$$\frac{}{() \Leftarrow () : ()} \quad (\text{TC_ZIP_I})$$

$$\frac{\Omega \Leftarrow \vec{x} : \vec{\psi}}{(\Omega, x : \psi) \Leftarrow (\vec{x}, x) : (\vec{\psi}, \psi)} \quad (\text{TC_ZIP_II})$$

**Zip quantified environment** $\boxed{\Theta \Leftarrow \vec{x} : \phi}$

$$\frac{\Omega \Leftarrow \vec{x} : \vec{\psi}}{[\Omega] \Leftarrow \vec{x} : [\vec{\psi}]} \quad (\text{TC_QZIP_I})$$

$$\frac{\Theta[I] \Leftarrow \vec{x} : \phi[I]}{\exists n\ \Theta[n] \Leftarrow \vec{x} : \exists n\ \phi[n]} \quad (\text{TC_QZIP_II})$$

**Init** $\boxed{\vec{x} : \psi = \omega}$

$$\frac{}{() : \psi = ()} \quad (\text{TC_INIT_I})$$

$$\frac{\vec{x} : \psi = \omega}{(\vec{x}, y) : \psi = (\omega, y : \psi)} \quad (\text{TC_INIT_II})$$

**Typecheck expression** $\boxed{\Gamma; \Omega \vdash e : \psi}$

$$\frac{x : \psi \in \Gamma}{\Gamma; \Omega \vdash x : \psi} \quad (\text{T_ENV_I})$$

$$\frac{x : \psi \in \Omega}{\Gamma; \Omega \vdash x : \psi} \quad (\text{T_ENV_II})$$

$$\frac{}{\Gamma; \Omega \vdash \star : \top} \quad (\text{T_TRUE})$$

$$\frac{}{\Gamma; \Omega \vdash 0 : \mathbf{nat}\,(0)} \quad (\text{T_ZERO})$$

$$\frac{\Gamma; \Omega \vdash q : \mathbf{nat}\,(i)}{\Gamma; \Omega \vdash \mathbf{s}(q) : \mathbf{nat}\,(succ(i))} \quad (\text{T_SUCC})$$

$$\frac{\vdash i_1 = i_2}{\Gamma; \Omega \vdash i_1 = i_2 : i_1 = i_2} \quad (\text{T_AX_I})$$

$$\frac{\vdash i_1 = i_2}{\Gamma; \Omega \vdash i_2 = i_1 : i_2 = i_1} \quad (\text{T_AX_II})$$

$$\frac{\Gamma; \Omega \vdash e : \psi[i_2] \qquad \Gamma; \Omega \vdash e' : i_1 = i_2}{\Gamma; \Omega \vdash e :> \{n/\psi[n]\}[e'] : \psi[i_1]} \quad (\text{T_EQUAL_E})$$

$$\frac{\Gamma; \Omega \vdash e : \mathbf{proc}\,\forall n\ \rho[n]}{\Gamma; \Omega \vdash e\{i\} : \mathbf{proc}\,\rho[i]} \quad (\text{T_PROC_INST})$$

$$\frac{\sim \exists n\ \phi[n] = \mathbf{proc}\,\forall n\ \rho[n] \qquad \Gamma; \Omega \vdash e : \mathbf{proc}\,\forall n\ \rho[n]}{\Gamma; \Omega \vdash e <: \{n/\phi[n]\}\{i\} : \mathbf{proc}\,\rho[i]} \quad (\text{T_CONT_INST})$$

$$\frac{\gamma = \vec{y} : \vec{\rho} \qquad \theta = \vec{z} : \phi \qquad \vec{z} : \top = \omega' \qquad \Gamma, \gamma = \Gamma' \qquad \Gamma'; \omega' \vdash s \triangleright \theta}{\Gamma; \Omega \vdash \mathbf{proc}\,[\gamma]\,\mathbf{out}\,\theta\{s\} : \mathbf{proc}\,([\vec{\rho}]\,\mathbf{out}\,\phi)} \quad (\text{T_PROC_DECL})$$

$$\frac{\forall I \cdot \Gamma; \Omega \vdash \mathbf{proc}\, h[I] : \mathbf{proc}\, \rho[I]}{\Gamma; \Omega \vdash \mathbf{proc}\, \forall n\ h[n] : \mathbf{proc}\, \forall n\ \rho[n]} \quad \text{(T_PROC_ABS)}$$

**Typecheck expressions** $\boxed{\Gamma; \Omega \vdash (\vec{e}) : (\vec{\psi})}$

$$\frac{}{\Gamma; \Omega \vdash () : ()} \quad \text{(T_EXPS_I)}$$

$$\frac{\Gamma; \Omega \vdash (\vec{e}) : (\vec{\psi}) \qquad \Gamma; \Omega \vdash e : \psi}{\Gamma; \Omega \vdash (\vec{e}, e) : (\vec{\psi}, \psi)} \quad \text{(T_EXPS_II)}$$

**Defined negation** $\boxed{\sim \phi = \psi}$

$$\frac{}{\sim [\vec{\psi}] = \sim (\vec{\psi})} \quad \text{(T_NEG_DEF_I)}$$

$$\frac{\forall N \cdot (\sim \phi[N] = \mathbf{proc}\, \rho[N])}{\sim \exists n\ \phi[n] = \mathbf{proc}\, \forall n\ \rho[n]} \quad \text{(T_NEG_DEF_II)}$$

**Typecheck sequence** $\boxed{\Gamma; \Omega \vdash s \triangleright \Theta}$

$$\frac{\Omega' \subset \Omega}{\Gamma; \Omega \vdash \varepsilon \triangleright [\Omega']} \quad \text{(T_EMPTY)}$$

$$\frac{\Gamma; \Omega \vdash s \triangleright \Theta[i]}{\Gamma; \Omega \vdash [i \in \exists n\ \Theta[n]] s \triangleright \exists n\ \Theta[n]} \quad \text{(T_WITNESS)}$$

$$\frac{\Gamma; \Omega \vdash e : i_1 = i_2 \qquad \Gamma; \Omega \vdash s \triangleright \Theta[i_2]}{\Gamma; \Omega \vdash s :> \exists n\ \Theta[n][e] \triangleright \Theta[i_1]} \quad \text{(T_SUBST)}$$

$$\frac{\Gamma; \Omega \vdash e : \psi \qquad \Gamma, y : \psi; \Omega \vdash s \triangleright \Theta}{\Gamma; \Omega \vdash \mathbf{cst}\, y = e; s \triangleright \Theta} \quad \text{(T_CST)}$$

$$\frac{\Gamma; \Omega \vdash e : \psi \qquad \Gamma; \Omega, y : \psi \vdash s \triangleright \Theta \qquad y \notin \Theta}{\Gamma; \Omega \vdash \mathbf{var}\, y := e; s \triangleright \Theta} \quad \text{(T_VAR)}$$

$$\frac{\Gamma; \Omega \vdash s \triangleright \theta \qquad \Gamma; \Omega[\![\theta]\!] \vdash s' \triangleright \Theta}{\Gamma; \Omega \vdash \{s\}_\theta; s' \triangleright \Theta} \quad \text{(T_BLOCK)}$$

$$\frac{\theta = \vec{x} : \phi \qquad \sim \phi = \psi \qquad \Gamma, y : \psi; \Omega \vdash s \triangleright \theta \qquad \Gamma; \Omega[\![\theta]\!] \vdash s' \triangleright \Theta}{\Gamma; \Omega \vdash y : \{s\}_\theta; s' \triangleright \Theta} \quad \text{(T_LABEL)}$$

$$\frac{\Gamma; \Omega \vdash e :\sim \vec{\psi} \qquad \Gamma; \Omega \vdash (\vec{e}) : (\vec{\psi}) \qquad \Gamma; \Omega[\![\theta]\!] \vdash s' \triangleright \Theta}{\Gamma; \Omega \vdash \mathbf{jump}\ (e, \vec{e})_\theta; s' \triangleright \Theta} \quad \text{(T_JUMP)}$$

$$\frac{y : \mathbf{nat}\,(i) \in \Omega \qquad \Gamma; \Omega[y : \mathbf{nat}\,(\mathit{succ}(i))] \vdash s \triangleright \Theta}{\Gamma; \Omega \vdash \mathbf{inc}\,(y); s \triangleright \Theta} \quad \text{(T_INC)}$$

$$\frac{y : \mathbf{nat}\,(i) \in \Omega \qquad \Gamma; \Omega[y : \mathbf{nat}\,(\mathit{pred}(i))] \vdash s \triangleright \Theta}{\Gamma; \Omega \vdash \mathbf{dec}\,(y); s \triangleright \Theta} \quad \text{(T_DEC)}$$

$$\frac{y : \psi \in \Omega \qquad \Gamma; \Omega \vdash e : \psi' \qquad \Gamma; \Omega[y : \psi'] \vdash s \triangleright \Theta}{\Gamma; \Omega \vdash y := e; s \triangleright \Theta} \quad \text{(T_ASSIGN)}$$

$$\frac{\omega[0] \subset \Omega \qquad \Gamma; \Omega \vdash e : \mathbf{nat}\,(i) \qquad \forall N \cdot \Gamma, y : \mathbf{nat}\,(N); \omega[N] \vdash s[N] \triangleright [\omega[\mathit{succ}(N)]] \qquad \Gamma; \Omega[\![\omega[i]]\!] \vdash s' \triangleright \Theta}{\Gamma; \Omega \vdash \mathbf{for}\ y : \mathbf{nat}(n) := 0\ \mathbf{until}\ e\ \{s[n]\}_{\omega[n]}; s' \triangleright \Theta} \quad \text{(T_FOR)}$$

$$\frac{\Gamma; \Omega \vdash e : \mathbf{proc}\,([\vec{\rho}]\,\mathbf{out}\,\phi) \qquad \Gamma; \Omega \vdash (\vec{e}) : (\vec{\rho}) \qquad \theta \Leftarrow \vec{z} : \phi \qquad \Gamma; \Omega[\![\theta]\!] \vdash s \triangleright \Theta}{\Gamma; \Omega \vdash e(\vec{e}; \vec{z}); s \triangleright \Theta} \quad \text{(T_CALL)}$$

**Typecheck sequence with updated environment** $\boxed{\Gamma; \Omega[\![\Theta]\!] \vdash s \triangleright \Theta'}$

$$\frac{\Gamma; \Omega[\![\Omega']\!] \vdash s \triangleright \Theta'}{\Gamma; \Omega[\![[\Omega']]\!] \vdash s \triangleright \Theta'} \quad \text{(TC_UPDATE_SEQ_I)}$$

$$\frac{\forall I \cdot \Gamma; \Omega[\![\Theta[I]]\!] \vdash s[I] \triangleright \Theta'}{\Gamma; \Omega[\![\exists n\ \Theta[n]]\!] \vdash ?n.s[n] \triangleright \Theta'} \quad \text{(TC_UPDATE_SEQ_II)}$$

## 3.3 Translation from ID to FD

**Type translation** $\boxed{\psi^\star = \varphi}$

$$\frac{}{\mathbf{nat}\,(i)^\star = \mathbf{nat}(i)} \quad \text{(TR_TYPE_NAT)}$$

$$\frac{}{x^\star = x} \quad \text{(TR_TYPE_VAR)}$$

$$\frac{}{\top^\star = \top} \quad \text{(TR_TYPE_TRUE)}$$

$$\frac{}{\bot^\star = \bot} \quad \text{(TR_TYPE_FALSE)}$$

$$\frac{}{(i_1 = i_2)^\star = i_1 = i_2} \quad \text{(TR_TYPE_EQUALS)}$$

$$\frac{(\rho)^\star = \varphi}{(\mathbf{proc}\,\rho)^\star = \varphi} \quad \text{(TR_TYPE_PROC)}$$

**Types translation** $\boxed{(\vec{\psi})^\star = (\vec{\varphi})}$

$$\frac{}{()^\star = ()} \quad \text{(TR_TYPES_I)}$$

$$\frac{(\vec{\psi})^\star = (\vec{\varphi}) \qquad \psi^\star = \varphi}{(\vec{\psi}, \psi)^\star = (\vec{\varphi}, \varphi)} \quad \text{(TR_TYPES_II)}$$

**Environment translation** $\boxed{(\gamma)^\star = (\vec{x}) : (\vec{\varphi})}$

$$\frac{}{()^\star = () : ()} \quad \text{(TR_ENV_I)}$$

$$\frac{(\gamma)^\star = (\vec{x}) : (\vec{\varphi}) \qquad \psi^\star = \varphi}{(\gamma, x : \psi)^\star = (\vec{x}, x) : (\vec{\varphi}, \varphi)} \quad \text{(TR_ENV_II)}$$

**Parametrized environment translation** $\boxed{(\theta)^\star = \vec{z} : \vec{\varphi}}$

$$\frac{\forall N \cdot (\gamma[N])^\star = (\vec{z}) : (\vec{\varphi}[N])}{(n \mapsto \gamma[n])^\star = \vec{z} : n \mapsto \vec{\varphi}[n]} \quad \text{(TR_ABS_ENV_I)}$$

**Parametrized type translation** $\boxed{(\vec{\psi})^\star = \varphi}$

$$\frac{\forall N \cdot \psi[N]^\star = \varphi[N]}{(\{n/\psi[n]\})^\star = \exists n\ \varphi[n]} \quad \text{(TR_ABS_TYPE_I)}$$

**Parametrized types translation** $\boxed{(\vec{\psi})^\star = (\vec{\varphi})}$

$$\frac{\forall N \cdot (\vec{\psi}[N])^\star = (\vec{\varphi}[N])}{(n \mapsto \vec{\psi}[n])^\star = (n \mapsto \vec{\varphi}[n])} \quad \text{(TR_ABS_TYPES_I)}$$

**Quantified types translation** $\boxed{(\phi)^\star = \varphi}$

$$\frac{(\vec{\psi})^\star = (\vec{\varphi})}{([\vec{\psi}])^\star = \langle\vec{\varphi}\rangle} \quad \text{(TR_QTYPES_I)}$$

$$\frac{\forall N \cdot (\phi[N])^\star = \varphi[N]}{(\exists n\ \phi[n])^\star = \exists n\ \varphi[n]} \quad \text{(TR_QTYPES_II)}$$

**Quantified types translation** $\boxed{(\theta)^\star = \langle \vec{x} \rangle : \varphi}$

$$\frac{(\gamma)^\star = (\vec{x}) : (\vec{\varphi})}{([\gamma])^\star = \langle \vec{x} \rangle : \langle \vec{\varphi} \rangle} \quad \text{(TR_QENV_I)}$$

$$\frac{\forall N \cdot (\theta[N])^\star = \langle \vec{x} \rangle : \varphi[N]}{(\exists n\ \theta[n])^\star = \langle \vec{x} \rangle : \exists n\ \varphi[n]} \quad \text{(TR_QENV_II)}$$

**Prototype translation** $\boxed{(\rho)^\star = \varphi}$

$$\frac{(\vec{\psi})^\star = (\vec{\varphi}) \qquad (\phi)^\star = \varphi'}{(([\vec{\psi}]\ \mathbf{out}\ \phi))^\star = \langle \vec{\varphi} \rangle \to \varphi'} \quad \text{(TR_PROTOTYPE_I)}$$

$$\frac{\forall N \cdot (\rho[N])^\star = \varphi[N]}{(\forall n\ \rho[n])^\star = \forall n\ \varphi[n]} \quad \text{(TR_PROTOTYPE_II)}$$

**Idents translation** $\boxed{(\vec{x})^\star = \vec{t}}$

$$\frac{}{()^\star = ()} \quad \text{(TR_IDENTS_I)}$$

$$\frac{(\vec{x})^\star = \vec{t}}{(\vec{x}, x)^\star = (\vec{t}, x)} \quad \text{(TR_IDENTS_II)}$$

**Number translation** $\boxed{q^\star = t}$

$$\frac{}{0^\star = 0} \quad \text{(TR_NUM_I)}$$

$$\frac{q^\star = t}{\mathbf{s}(q)^\star = \mathbf{succ}\,(t)} \quad \text{(TR_NUM_II)}$$

**Header translation** $\boxed{(h)^\star = t}$

$$\frac{\theta = \vec{z} : \phi \qquad (s)^\star_{\vec{z}} = t \qquad (\gamma)^\star = (\vec{x}) : (\vec{\varphi})}{([\gamma]\ \mathbf{out}\ \theta\{s\})^\star = \mathbf{fn}\,(\vec{x} : \vec{\varphi}) \Rightarrow t} \quad \text{(TR_HEADER_I)}$$

$$\frac{\forall N \cdot (h[N])^\star = t[N]}{(\forall n\ h[n])^\star = \lambda n.t[n]} \quad \text{(TR_HEADER_II)}$$

**Expression translation** $\boxed{(e)^\star = t}$

$$\frac{q^\star = t}{(q)^\star = t} \quad \text{(TR_EXP_NUM)}$$

$$\frac{}{(x)^\star = x} \quad \text{(TR_EXP_VAR)}$$

$$\frac{}{(\star)^\star = \langle\rangle} \quad \text{(TR_EXP_STAR)}$$

$$\frac{}{(i_1 = i_2)^\star = i_1 = i_2} \quad \text{(TR_EXP_AXIOM)}$$

$$\frac{(h)^\star = t}{(\mathbf{proc}\ h)^\star = t} \quad \text{(TR_EXP_PROC)}$$

$$\frac{(e)^\star = t}{(e\{i\})^\star = t\{i\}} \quad \text{(TR_EXP_INST)}$$

$$\frac{(e)^\star = t \quad \forall N \cdot (\phi[N])^\star = \varphi[N]}{(e <: \{n/\phi[n]\}\{i\})^\star = \textbf{fn}\, v_1 : \varphi[i] \Rightarrow (t\,\langle i, v_1 : \exists n\ \varphi[n]\rangle)} \quad \text{(TR_EXP_INST')}$$

$$\frac{(e)^\star = t \quad (e')^\star = t' \quad (\{n/\psi[n]\})^\star = \varphi}{(e :> \{n/\psi[n]\}[e'])^\star = t :> \varphi[t']} \quad \text{(TR_EXP_SUBST)}$$

**Expressions translation** $\boxed{(\vec{e})^\star = (\vec{t})}$

$$\frac{}{()^\star = ()} \quad \text{(TR_EXPS_I)}$$

$$\frac{(\vec{e})^\star = (\vec{t}) \quad (e)^\star = t}{(\vec{e}, e)^\star = (\vec{t}, t)} \quad \text{(TR_EXPS_II)}$$

**Sequence translation** $\boxed{(s)^\star_{\vec{x}} = t}$

$$\frac{(\vec{x})^\star = \vec{t}}{()^\star_{\vec{x}} = \langle \vec{t} \rangle} \quad \text{(TR_SEQ_EMPTY)}$$

$$\frac{(e)^\star = t \quad (s)^\star_{\vec{x}} = t'}{(\textbf{var}\, x := e; s)^\star_{\vec{x}} = \textbf{let}\, x = t\, \textbf{in}\, t'} \quad \text{(TR_SEQ_VAR)}$$

$$\frac{(e)^\star = t \quad (s)^\star_{\vec{x}} = t'}{(\textbf{cst}\, x = e; s)^\star_{\vec{x}} = \textbf{let}\, x = t\, \textbf{in}\, t'} \quad \text{(TR_SEQ_CST)}$$

$$\frac{(e)^\star = t \quad (s)^\star_{\vec{x}} = t'}{(x := e; s)^\star_{\vec{x}} = \textbf{let}\, x = t\, \textbf{in}\, t'} \quad \text{(TR_SEQ_ASSIGN)}$$

$$\frac{(s)^\star_{\vec{x}} = t}{(\textbf{inc}\,(x); s)^\star_{\vec{x}} = \textbf{let}\, x = \textbf{succ}\,(x)\, \textbf{in}\, t} \quad \text{(TR_SEQ_INC)}$$

$$\frac{(s)^\star_{\vec{x}} = t}{(\textbf{dec}\,(x); s)^\star_{\vec{x}} = \textbf{let}\, x = \textbf{pred}\,(x)\, \textbf{in}\, t} \quad \text{(TR_SEQ_DEC)}$$

$$\frac{(e)^\star = t \quad (\vec{e})^\star = (\vec{u}) \quad (s)^\star_{\vec{x}} = t'}{(e(\vec{e}; \vec{z}); s)^\star_{\vec{x}} = \textbf{let}\, \langle \vec{z} \rangle = t\, \langle \vec{u} \rangle\, \textbf{in}\, t'} \quad \text{(TR_SEQ_CALL)}$$

$$\frac{\theta = \vec{z} : \phi \quad (s_1)^\star_{\vec{z}} = t_1 \quad (s_2)^\star_{\vec{x}} = t_2}{(\{s_1\}_\theta; s_2)^\star_{\vec{x}} = \textbf{let}\, \langle \vec{z} \rangle = t_1\, \textbf{in}\, t_2} \quad \text{(TR_SEQ_BLOCK)}$$

$$\frac{(n \mapsto \omega[n])^\star = \vec{z} : n \mapsto \vec{\varphi}[n] \quad (e)^\star = u' \quad (\vec{z})^\star = \vec{u} \quad (n \mapsto s_1[n])^\star_{\vec{z}} = n \mapsto t[n] \quad (s_2)^\star_{\vec{x}} = t'}{(\textbf{for}\ y : \textbf{nat}(n) := 0\ \textbf{until}\ e\ \{s_1[n]\}_{\omega[n]}; s_2)^\star_{\vec{x}} = \textbf{let}\, \langle \vec{z} \rangle = \textbf{rec}(u', \langle \vec{u} \rangle, \lambda n.\textbf{fn}\, y : \textbf{nat}(n) \Rightarrow \textbf{fn}\,(\vec{z} : \vec{\varphi}[n]) \Rightarrow t[n])\, \textbf{in}\, t'} \quad \text{(TR_SEQ_FOR)}$$

$$\frac{\forall N \cdot (s[N])^\star_{\vec{x}} = t[N]}{(?n.s[n])^\star_{\vec{x}} = ?n.t[n]} \quad \text{(TR_SEQ_ANY)}$$

$$\frac{(s)^\star_{\vec{x}} = t \quad (\theta)^\star = \langle \vec{z} \rangle : \varphi}{([i \in \theta]s)^\star_{\vec{x}} = \langle i, t : \varphi \rangle} \quad \text{(TR_SEQ_WITNESS)}$$

$$\frac{(s)^\star_{\vec{x}} = t \quad (e)^\star = u \quad (\theta)^\star = \langle \vec{z} \rangle : \varphi}{(s :> \theta[e])^\star_{\vec{x}} = t :> \varphi[u]} \quad \text{(TR_SEQ_SUBST)}$$

$$\frac{(e)^\star = t \quad (\vec{e})^\star = (\vec{u}) \quad (s)^\star_{\vec{x}} = t' \quad (\theta)^\star = \langle \vec{z} \rangle : \varphi}{(\textbf{jump}\ (e, \vec{e})_\theta; s)^\star_{\vec{x}} = \textbf{let}\, \langle \vec{z} \rangle = \textbf{throw}_\varphi\ t\ \langle \vec{u} \rangle\, \textbf{in}\, t'} \quad \text{(TR_SEQ_JUMP)}$$

$$\frac{(\theta)^\star = \langle \vec{z} \rangle : \varphi \quad (s)^\star_{\vec{z}} = t \quad (s')^\star_{\vec{x}} = t'}{(y : \{s\}_\theta; s')^\star_{\vec{x}} = \textbf{let}\, \langle \vec{z} \rangle = \textbf{callcc}\,(\textbf{fn}\, y : \neg\varphi \Rightarrow t)\, \textbf{in}\, t'} \quad \text{(TR_SEQ_LABEL)}$$

**Loop body translation** $\boxed{(b)^{\star}_{\vec{x}} = t}$

$$\frac{\forall N \cdot (s[N])^{\star}_{\vec{x}} = t[N]}{(n \mapsto s[n])^{\star}_{\vec{x}} = n \mapsto t[n]} \qquad \text{(TR_BODY_ABS)}$$